\documentclass{aa}
\usepackage[varg]{txfonts}
\usepackage{booktabs,graphicx,siunitx,xspace,natbib}
\DeclareSIUnit \parsec {pc}
\DeclareSIUnit \pc     {\parsec}
\DeclareSIUnit \kpc    {\kilo \parsec}
\DeclareSIUnit \msun   {\ensuremath{M_\sun}}
\DeclareSIUnit \year   {a}
\newcommand{\about}{$\simeq$}

\newcommand{\Co}{$^{56}$Co\xspace} %x

\newcommand{\Ni}{$^{56}$Ni\xspace}

\newcommand{\Msol}{M\ensuremath{_\odot}\xspace}

\begin{document}

\title{SN2014J gamma-rays from the $^{56}$Ni decay chain}

\author{
  Roland Diehl         \inst{\ref{inst:mpe},\ref{inst:xcu}} \and
  Thomas Siegert   \inst{\ref{inst:mpe}} \and
  Wolfgang Hillebrandt        \inst{\ref{inst:mpa},\ref{inst:xcu}} \and
  Martin Krause        \inst{\ref{inst:xcu},\ref{inst:mpe}} \and
  Jochen Greiner       \inst{\ref{inst:mpe},\ref{inst:xcu}} \and
    Keiichi Maeda       \inst{\ref{inst:kyoto}, \ref{inst:tokyo}} \and
      Friedrich K. R\"opke        \inst{\ref{inst:wue}} \and
      Stuart A. Sim        \inst{\ref{inst:belfast}} \and
    Wei Wang       \inst{\ref{inst:naoc},\ref{inst:xcu}, \ref{inst:mpe}} \and
     Xiaoling Zhang       \inst{\ref{inst:mpe}} 
     %\and
  %{\it et al.}
}

\institute{
  Max-Planck-Institut f\"ur extraterrestrische Physik,
  D-85741 Garching, Germany
  \label{inst:mpe}
    \email{rod@mpe.mpg.de}
 \and
  Excellence Cluster Universe, Boltzmannstr. 2, D-85748, Garching, Germany
  \label{inst:xcu}
    \and
  Max-Planck-Institut f\"ur Astrophysik,
  D-85741 Garching, Germany
  \label{inst:mpa}
    \and
  Dept. of Astronomy, Kyoto University, Kitashirakawa-Oiwake-cho, Sakyo-ku, Kyoto 606-8502, Japan
   \label{inst:kyoto}  \and 
  Kavli Institute for Physics and Mathematics of the Universe (WPI), University of Tokyo, 5-1-5 Kashiwanoha, Kashiwa, Chiba 277-8583, Japan
   \label{inst:tokyo}
  \and
  Institut f\"ur Theoretische Physik und Astrophysik, Universit\"at W\"urzburg, Emil-Fischer-Str. 31, 97074 W\"urzburg, Germany
  \label{inst:wue}
  \and
  Astrophysics Research Centre, School of Mathematics and Physics, Queens University, University Road, Belfast, BT7 1NN, UK
  \label{inst:belfast}
  \and
  National Astronomical Observatories, Chinese Academy of Sciences,   Beijing 100012, China
  \label{inst:naoc}
  }

\date{Received 15 September 2014 / Accepted 16 December 2014}

\abstract
% context heading (can be left empty)
{The detection and measurement of gamma-ray lines from the decay chain of $^{56}$Ni provides
 unique information about the explosion in supernovae. SN2014J at 3.3~Mpc is a sufficiently nearby supernova of type Ia
   so that such measurements have been feasible with the gamma-ray spectrometer SPI on ESA's INTEGRAL gamma-ray observatory.}
% aims heading
{The $^{56}$Ni freshly-produced in the supernova is understood to power the optical light curve, as it emits gamma-rays upon its radioactive decay first to \Co and then to $^{56}$Fe. Gamma-ray lines from $^{56}$Co decay are expected to become directly visible through the overlying white dwarf material several weeks after the explosion,  as they progressively penetrate the overlying material of the supernova envelope, diluted as it expands. The lines are expected to be Doppler-shifted or broadened from the kinematics of the $^{56}$Ni ejecta.
We aim to exploit high-resolution gamma-ray spectroscopy with the SPI spectrometer on INTEGRAL towards
constraining the $^{56}$Ni distribution and kinematics in this supernova.}
% methods heading
{We use the observations with the SPI spectrometer on INTEGRAL 
  together with an improved instrumental background method.}
% results heading
{We detect the two main lines from \Co decay at 847 and 1238 keV, significantly Doppler-broadened, and at intensities (3.65$\pm$1.21)~10$^{-4}$ and (2.27$\pm$0.69)~10$^{-4}$~ph~cm$^{-2}$s$^{-1}$, respectively, at brightness maximum. We measure their rise towards a maximum
 after about 60--100 days and decline thereafter. 
 %We note that the identification of the \Co lines is difficult, as we try to exploit both temporal and spectral resolution of our measurement.
The intensity ratio of the two lines is found consistent with expectations from \Co decay (0.62$\pm$0.28 at brightness maximum, expected is 0.68). 
 We find that the broad lines seen in the late, gamma-ray transparent phase are not representative for the early gamma-ray emission, 
and rather notice the emission spectrum to be complex and irregular until the supernova is fully transparent to gamma-rays, with progressive uncovering of the bulk of \Ni. 
 We infer that the explosion morphology is not spherically symmetric, both in the distribution of \Ni  and of the unburnt material which occults the \Co emission. Comparing light curves from different plausible models, the resulting \Ni mass is determined as 0.49$\pm$0.09~\Msol. 
 }
% conclusions heading (can be left empty)
{}


\maketitle

% ----------------------------------------------------------------------
%
\section{Introduction}
\label{sec:intro}

Supernovae of type Ia are generally understood to arise from thermonuclear disruption of a CO white dwarf in a binary system, caused by the rapid nuclear energy release from carbon fusion ignited in the central region of the white dwarf \citep{Hillebrandt:2013aa}. Different scenarios are discussed on how the explosion may be initiated: Either the accretion of material from the companion star causes the white dwarf to reach the Chandrasekhar-mass stability limit, or an external event on the white dwarf such as a major accretion event or a nuclear explosion on the surface makes the white dwarf interior unstable towards runaway nuclear carbon fusion. Once ignited, nuclear fusion at high densities processes the white dwarf material to iron group nuclei, which are the most stable configuration of nuclear matter, with radioactive \Ni being a major product of such explosive supernova nucleosynthesis \citep{Nomoto:1997aa}. As the nuclear flame rushes through the star, an explosion is initiated, and nuclear burning then competes with expansion of the material, resulting in some outer parts of the white dwarf not being burnt towards iron group nuclei, but only to intermediate-mass elements, or even left unburnt as carbon and oxygen mainly \citep{Mazzali:2007aa}. Typically, it is expected that about 0.5 \Msol of \Ni are thus produced and embedded in about 0.5--0.9~\Msol of other material \citep{Mazzali:2007aa,Stritzinger:2006aa}. As the supernova expands, more and more of the \Ni gamma-rays from radioactive decay thus should be able to leave the source region where the decay occurs, and be observable with gamma-ray telescopes \citep{Isern:2008aa}.

\Ni radioactive decay occurs with a first decay to \Co after $\tau\sim$~8 days, i.e. when the supernova is expected to be still opaque to even gamma-rays at MeV energies, converting this radioactivity energy into emission at lower energy photons \citep{Hoeflich:1998ab}. When the second decay stage from \Co to $^{56}$Fe at $\tau\sim$~111 days occurs, producing gamma-rays at 846.77 and 1238.29 keV, the supernova envelope begins to be transparent. After a few months, full transparency to gamma-rays will be reached, and the radioactive decay causes the intensity of characteristic gamma-rays to fade away. Between initial gamma-ray leakage and the stage of full transparency, however, the brightness of \Co decay gamma-rays is determined by the amount of available \Ni and how it is distributed within the expanding supernova. This is the phase, where the rise and fall of the gamma-ray line intensity provides unique information on the type of explosion and the structure of the supernova. 

SN2014J was discovered on January 22, 2014 \citep{Fossey:2014aa}, and recognised as a type Ia explosion from early spectra \citep{Cao:2014aa}. It occurred in the nearby starburst galaxy M82 at \about~3.3 Mpc distance \citep{Foley:2014}, and is seen only through major interstellar material in the host galaxy along the line-of-sight, causing rather large reddening \citep{Goobar:2014aa} with corresponding difficulties to uncover the proprietary supernova brightness and low-energy spectra at the desired precision. The explosion date, therefore, has been inferred as 14 January, UT 14.75, with 0.21~d uncertainty \citep{Zheng:2014aa}. The light curve reached its blue-band maximum about 20 days after the explosion \citep{Goobar:2014aa}. Our detection of \Ni gamma-rays early on \citep{Diehl:2014} also show that the initiation of the supernova explosion may have been unusual and from a primary nuclear explosion near the surface of the white dwarf, causing additional uncertainty on the structure of the outer supernova envelope. It is therefore of great interest to study the evolution of the gamma-rays from \Co decay, as they trace the release of \Ni radioactive energy and its occultation. This should complement data obtained from the UV-optical-IR photosphere further outside, which reveals photospheric material composition through a variety of atomic absorption lines, and can be interpreted using radiation transport calculations \citep{Dessart:2014,Dessart:2014a}.

INTEGRAL started observing SN2014J on 31 January, and kept monitoring  the supernova over about 7 Msec until 26 June, when visibility constraints terminated this special opportunity. This covers SN2014J emission during days 16.3 -- 164.0 after the explosion. Here we describe our analysis of data from the SPI instrument on INTEGRAL, which is unique with keV energy resolution to reveal details of the shape of the characteristic gamma-ray lines from the \Ni decay chain, thus providing kinematic constraints, in addition to tracing the evolution of brightness. 

% ----------------------------------------------------------------------
%
\section{Data and their Analysis}
\label{sec:spi-analysis}
% ----------------------------------------------------------------------
%

The INTEGRAL space gamma-ray observatory of ESA \citep{Winkler:2003} carries the gamma-ray spectrometer instrument 'SPI' as one of its two main instruments \citep{Vedrenne:2003}. Both INTEGRAL main telescopes utilize the \emph{coded mask} technique for imaging gamma-ray sources. In this technique, a mask with occulting tungsten blocks and holes in the field of view of the gamma-ray camera imprints a shadowgram of a celestial source signal in the 19-element detector plane. The SPI spectrometer features a camera consisting of 19 high-resolution Ge detectors, which provides a spectrum of celestial gamma-rays attributed to their source through coded-mask shadowgrams, above a large instrumental background. SPI data consist of energy-binned spectra for each of the 15 Ge detectors of the SPI telescope camera which were operational during our observations of SN2014J. 

For our analysis, we used exposures of SN2014J accumulated over a special campaign \citep{Kuulkers:2014} of the INTEGRAL
mission, during orbit numbers 1380 to 1428, with one major gap between 23 April and 27 May.   The main exposure window of 2.8~Msec was placed in the rising part of the expected gamma-ray line emission, with an additional 1.4~Msec exposure at late times when occultation of gamma-rays should be a minor effect only, and the total \Co decay would be observed. 
After data selections to suppress contaminations, e.g. from solar flare events, our dataset consists of 4.2 Msec of exposure, spread over 1816 telescope pointings.
Exposures are taken in typically \SI{3000}{\second} pointings of the telescope towards this sky region, then moving the telescope pointing by 2.1 degrees to shift the shadowgram of the source in the detector plane. A regular pattern of telescope pointings of a 5 by 5 rectangle around the direction of the supernova itself comprises one such cycle of the dithering exposure pattern. 
The supernova always was in the telescope field of view, with its \about 30 degree opening angle, and variations of sensitivity due to different aspect angles were $<$10\%.

Our analysis method is based on a comparison of measured data to models, performed in a data space consisting of the counts per energy bin measured in each of SPI's detectors for each single exposure of the complete observation.
We describe data $d_k$ per energy bin $k$  in general terms as a linear combination of the sky model
components $M_{ij}$, to which the instrument response matrix $R_{jk}$
is applied, and the background components $B_{jk}$, with parameters $\theta_i$ for $N_I$ sky and $N_B$ background components:
\begin{equation}
d_k = \sum_j R_{jk} \sum_{i = 1}^{N_\mathrm{I}} \theta_i M_{ij} +
\sum_{i = N_\mathrm{I} + 1}^{N_\mathrm{I} + N_\mathrm{B}} \theta_i
B_{jk}\label{eq:model-fit}
\end{equation}
Generally, in our spectroscopy analysis we fit the intensity scaling factor of a model of the sky
intensity distribution plus a set of scaling factors for a model of the instrumental
background to data in energy bins covering the spectral range of interest. 
The response matrix encodes the shadowgram effects, i.e., how the occultation by the coded mask affects visibility of the source direction from each of the Ge detectors of the camera. In our case, we use a single sky component for the SN2014J point source ($N_I=1$), and a single background model ($N_B=1$). The latter is derived by a detailed spectroscopic assessment of longterm background and detector behaviour. 

Our treatment of instrumental background follows a new approach, which accounts for the physical nature of instrumental background lines and of detector-specific spectral responses, combining data across a broader range of energy and time periods suitably to build a self-consistent description of spectral detector response and background and their variations. Continuum and line backgrounds are treated separately, individual detector responses and their degradations are determined from a combination of spectral lines and their longterm behaviour. Recognising physical processes which cause characteristic instrumental lines has been part of the validation of the background determination.This method is described in detail in \citet{Diehl:2014}, and was applied for the \Ni decay lines from SN2014J successfully; we refer to the Supplementary Information therein for details of the data analysis method and its validation.

Our entire dataset for SN2014J consists  of 27240 spectra accumulated from single-detector events in 0.5~keV energy bins. 
We chose to fit five parameters in this analysis: one intensity scaling for our complete background model, and an intensity amplitude for the SN2014J signal itself for four different epochs. This compromise attempts to account for SN201J gamma-ray variation on the scale of \about~a month, as expected from models \citep[see, e.g.][]{The:2014}, avoiding to prescribe a model/assumption about the rise and fall of the gamma-ray line intensity, beyond a four-element gamma-ray light curve. We also employ an analysis of eleven different time epochs, when we want to investigate evolutions of spectral features on shorter times of \about~two weeks. Alternatively, we also fit brightness evolutions from a set of candidate models to our measurements, with normalisations of the respective SN2014J light curve model and of instrumental background.

% ----------------------------------------------------------------------
\begin{figure}
  \centering
  \includegraphics[width=\linewidth]{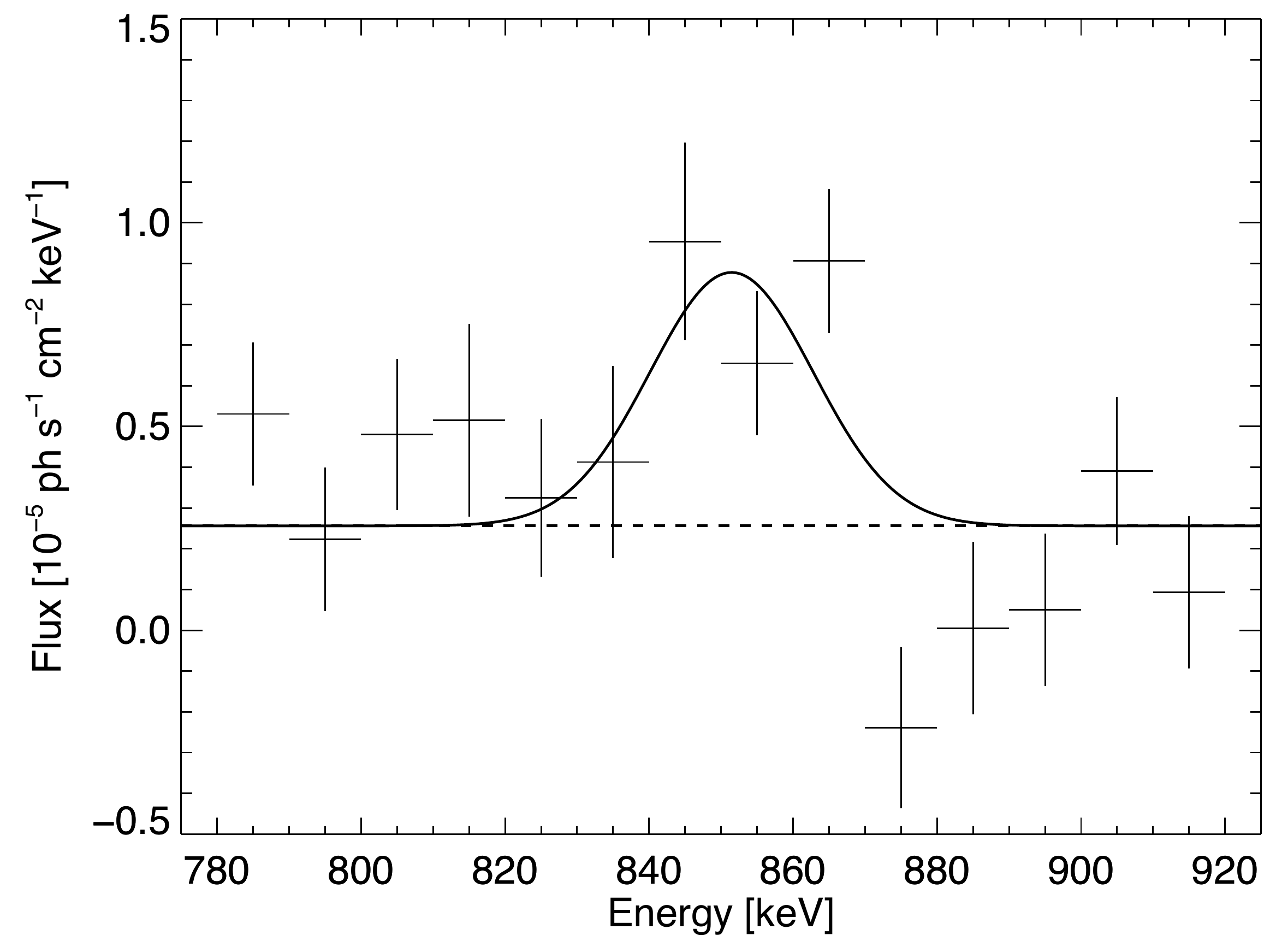}
  \includegraphics[width=\linewidth]{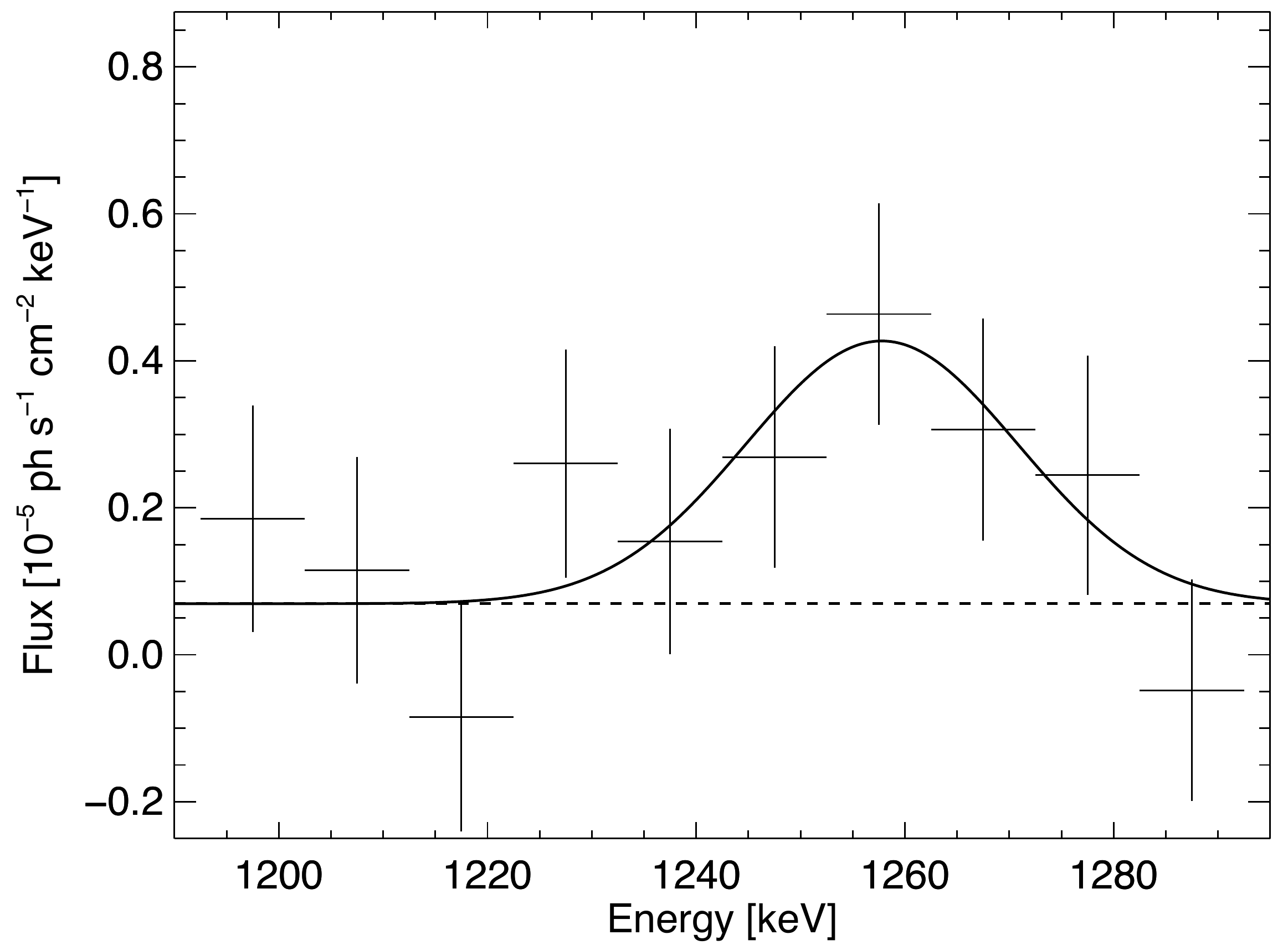}
   \caption{SN2014J spectrum near the 847 keV line ({\it above}) and near the 1238 keV line ({\it below}) as expected from \Co decay. These spectra are determined in energy bins  of width 10 keV over the entire observing period; the source intensity is fitted at four independent epochs. For illustration, fitted Gaussians indicate the detection of broadened lines near the \Co gamma-ray line energies.}
  \label{fig:SPI-spectrum-847-1238}
\end{figure}
% ----------------------------------------------------------------------
% ----------------------------------------------------------------------
%
\section{Results and Discussion}
\label{sec:results}

\subsection{Detection of \Co lines}

We determine the spectrum of the gamma-ray signal from SN2014J in two energy bands around the expected \Co decay lines, which have energies at rest of 846.77 and 1238.29~keV, the higher-energy line having 68\% of the 847~keV line intensity due to the branching ratio of the nuclear de-excitation. The energy bands chosen are 780 to 920 keV (around the 847~keV line) and 1190 to 1290~keV (around the 1238~keV line). We expect Doppler shift and broadening effects, which would be on the order of 15~keV (21~keV) for 5000 km~s$^{-1}$ velocity along the line of sight. 

Fig.~\ref{fig:SPI-spectrum-847-1238} shows the spectrum for SN2014J, which was derived from the entire observations set covering days 17 -- 164 after the supernova explosion, for the two strongest lines emitted in radioactive decay of \Co. These integrated time-averaged spectra were derived from fitting a source at the position of SN2014J in four independent epochs (see below for details), thus allowing for time variability of the flux, as expected. 
%The epochs we used are post-explosion days 16.3--41.3, 41.3--66.3, 66.3-99.1, and 134.8-164.0. 

% ----------------------------------------------------------------------
\begin{figure}
  \centering
  \includegraphics[width=\linewidth]{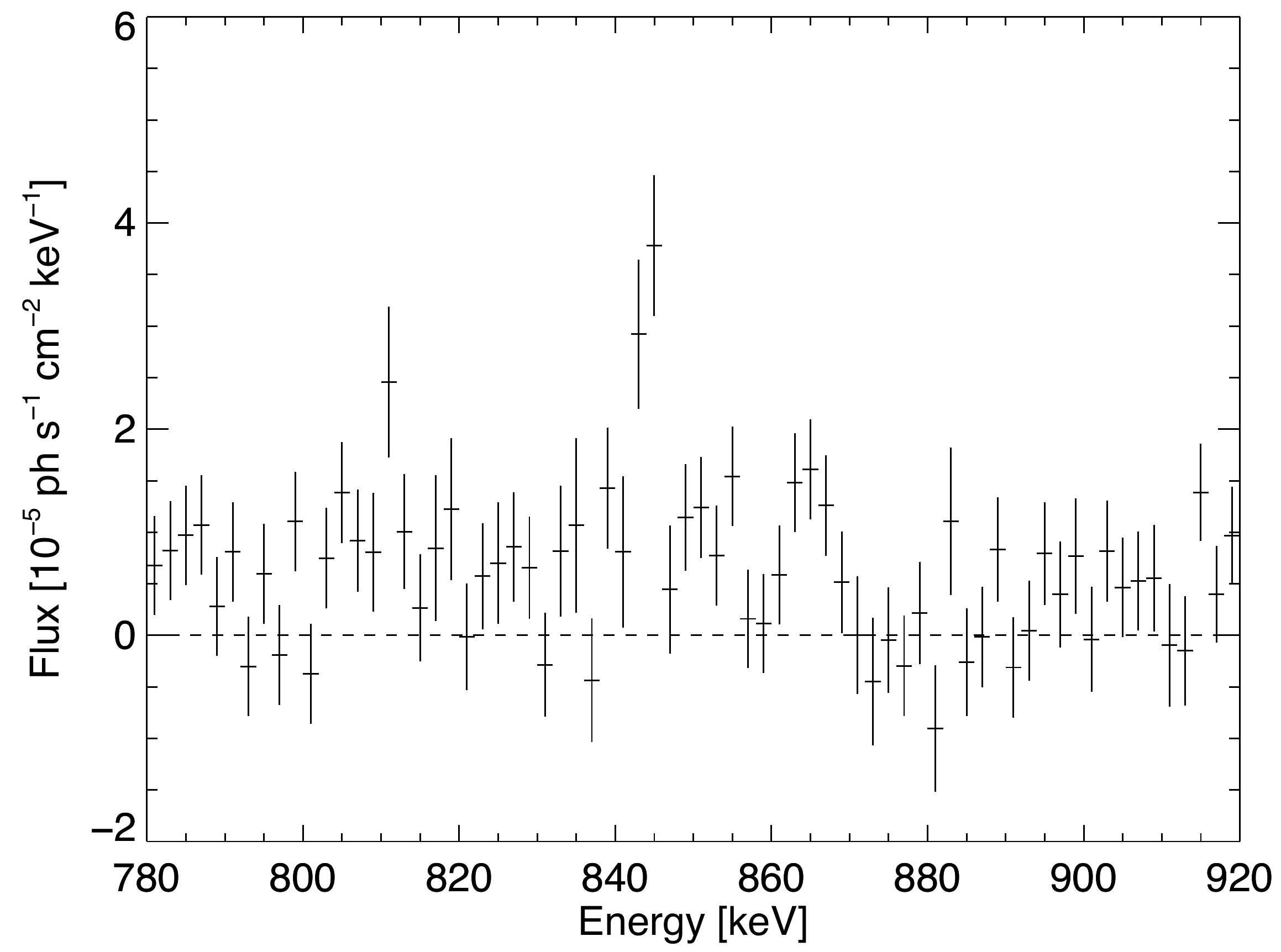}
   \caption{SN2014J spectrum near the 847 keV line as shown in Fig.~\ref{fig:SPI-spectrum-847-1238}, but here the analysis was performed  in 2~keV energy bins, which corresponds to the instrumental resolution, and for eleven epochs separately, before summed up. Apparently, a single broad Gaussian does not capture the line shape properly across the rise, peak, and fall of the gamma-ray emission.}
  \label{fig:SPI-spectrum-847-1}
\end{figure}
% ----------------------------------------------------------------------

These spectra show significant emission from SN2014J, overall dominated by broadened lines centered near 847 and 1238 keV, as expected. Gaussian profiles as shown were fitted together with an offset accounting for possible continuum. The flux error bars shown per data point were determined from propagating Poissonian uncertainties through our maximum-likelihood fitting of instrument and background model to the measured dataset; horizontal bars indicate the 10~keV wide energy bins. 
Overall, the significance of line emission detected from SN2014J in these two energy bands is 9.5 and 3.1~$\sigma$, for the 780 to 920 keV and 1190 to 1290~keV bands, respectively.  
%This spectrum is practically identical to the spectrum derived by another analysis from a large subset of the same data  \citep{Churazov:2014}.
Generally, we detect characteristic \Co gamma-ray line emission in agreement with first-order models of SNIa explosions, and also consistent with results reported by  \citet{Churazov:2014}. The lines are modestly broadened and somewhat offset as the \Ni produced initially in the explosion is partly occulted behind envelope material, uncovering \Ni more and more with time and ejecta dilution. The detection has an overall significance of \about~10~$\sigma$, limited from the instrumental background at \about hundred times higher count rate, and Poissonian statistics.

But one expects that the gamma-ray lines from \Co decay are gradually emerging from the supernova, as the overlying material becomes transparent to the gamma-rays with successive supernova expansion \citep{Isern:1997aa}. At earlier times, spectrum and intensity of gamma-ray emission from the primary \Ni and \Co energy source may appear different from the late, gamma-ray transparent phase, where \Co decay and ejecta kinematics determines the gamma-ray signal.
When we analyze the same dataset of our observations in separate epochs and in finer energy bins for the 780--920~keV band, we obtain a cumulative spectrum shown in Fig.~\ref{fig:SPI-spectrum-847-1}. Here we separated our observations in eleven time epochs (details see below), and 2 keV wide energy bins of analysis, close to the instrumental line width of 2.3~keV. This spectrum reveals that the fitted Gaussian profiles in Fig.~\ref{fig:SPI-spectrum-847-1238} may not capture the actual line emission as it evolves.%, and rather, multiple emission components are indicated, which may originate also from different times during the evolution.
%Therefore, we abandon interpretation of the cumulative spectra of Fig.~\ref{fig:SPI-spectrum-847-1238}, determined in the entire phase of gradually increasing gamma-ray transparency of the supernova, and rather inspect the spectra as they evolve during our observations. 

% ----------------------------------------------------------------------
\begin{figure}
  \centering
  \includegraphics[width=\linewidth]{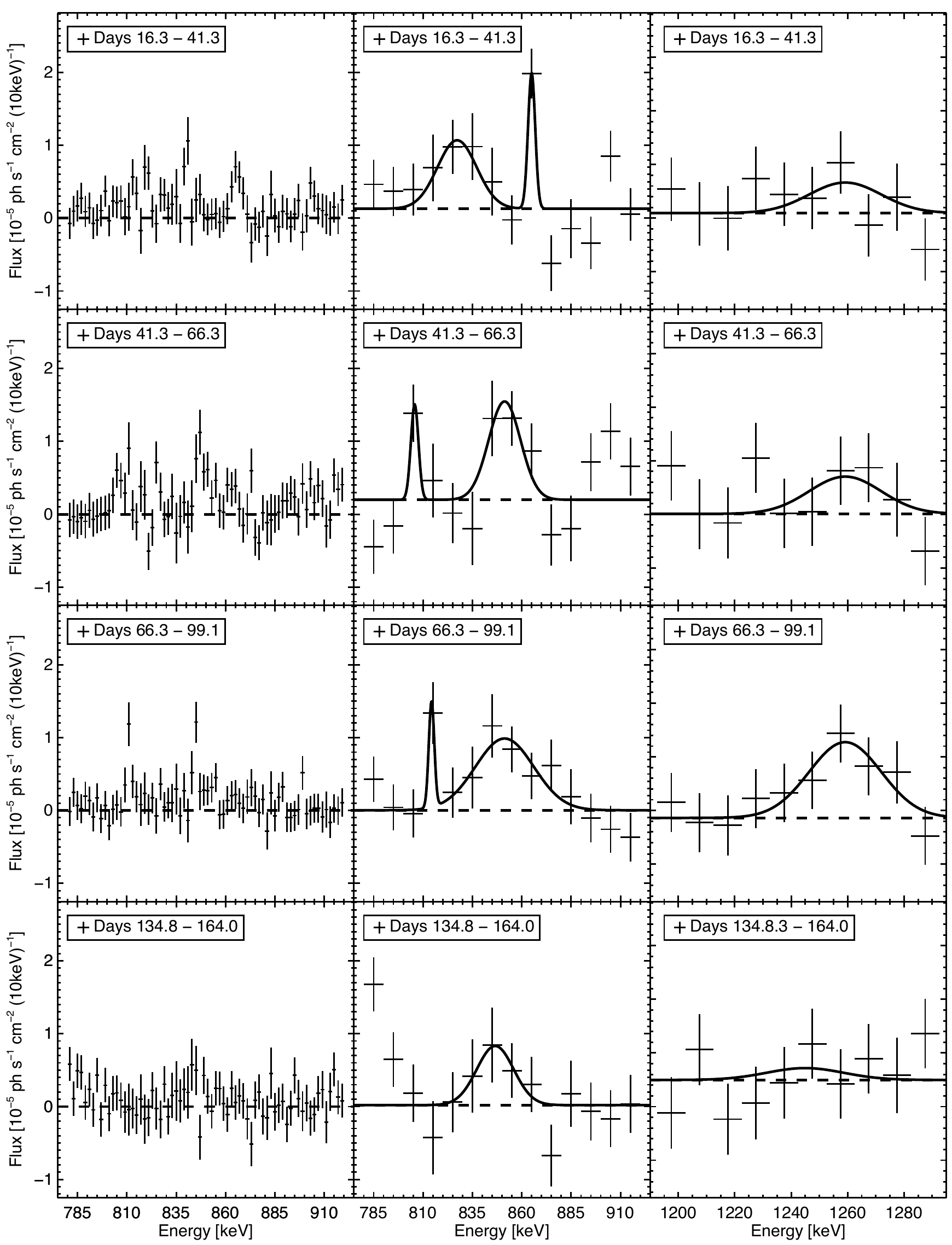}
   \caption{SN2014J signal intensity variations for the 847 keV line ({\it center}) and  the 1238 keV line ({\it right}) as seen in the four epochs of our observations, and analyzed with 10 keV energy bins. The 1238 keV fluxes have been scaled by the \Co decay branching ratio of 0.68 for equal-intensity appearance. Clear and significant emission is seen in the lower energy band ({\it left and center}) through a dominating broad line attributed to 847~keV emission, the emission in the high-energy band in the 1238~keV line appears consistent and weaker, as expected from the branching ratio of 0.68  ({\it right}). Fitted line details are discussed in the text.
   For the 847 keV line, in addition a high-spectral resolution analysis is shown at 2 keV energy bin width ({\it left}), confirming the irregular, non-broad-Gaussian features in more detail.  }
  \label{fig:SPI-spectra-set-847-1238}
\end{figure}
% ----------------------------------------------------------------------

\subsection{Temporal and spectral variations}

\subsubsection{Towards a light curve -- four epochs}

In a first approach towards time-resolved gamma-ray emission, we subdivide our observations into four epochs of post-explosion days 16.3--41.3 (epoch 1), 41.3--66.3 (2), 66.3-99.1 (3), and 134.8-164.0 (epoch 4). Here the first three epochs fall into the phase of gradually rising emission, while after the observation gap between days 100 and 134, the fourth epoch should capture the \Co emission in a rather transparent supernova. We again use 10~keV wide energy bins for the analysis, for a high signal-to-noise ratio per bin, which should be adequate as lines are expected to be kinematically broadened (see Fig.~\ref{fig:SPI-spectrum-847-1238}). 
Fig.~\ref{fig:SPI-spectra-set-847-1238} shows the individual spectra measured at the four different epochs, together with Gaussians fitted to the major features.

At the brightest epoch (3), \Co decay gamma-ray line intensities are found as (3.65$\pm$1.21)~10$^{-4}$ and (2.27$\pm$0.69)~10$^{-4}$~ph~cm$^{-2}$s$^{-1}$ in the 847 and 1238~keV lines, respectively. 
%We find an intensity ratio of the two lines as (0.62~$\pm$0.28), expected from \Co decay is a ratio of 0.68. This indicates that at this time the supernova is sufficiently diluted such that the \Co decay lines can escape from the bulk of \Ni. %% this is presented now below%%
From the flux measured at this time, we estimate a \Ni mass of 0.50$\pm$0.12~\Msol, assuming supernova transparency (see however below, for an alternative \Ni mass estimate).

Now we investigate the time evolution of the detected spectral features in more detail. We aim to trace the intensity variations in time steps that are able to discriminate among different models for the explosion, although we may expect that fainter emission before and after maximum flux will result in larger uncertainty of the flux determined at this time.

%Here we analyze separately each line and epoch, first determinig the SN2014J spectrum, and then modeling it by fitting a Gaussian. 

% Table of Line fitting results
\begin{table}[htbp]
   \centering \small
   \begin{tabular}{@{} ccccccc @{}} % Column formatting, @{} suppresses leading/trailing space
      \toprule
     \multicolumn{7}  {c}  {847 keV line}   {1238 keV line
     } \\
      \cmidrule(r){2-4} % Partial rule. (r) trims the line a little bit on the right; (l) & (lr) also possible
      \cmidrule(l){5-7} % Partial rule. (r) trims the line a little bit on the right; (l) & (lr) also possible
      time  & flux & center    & width & flux & center    & width  \\
      \midrule
        16.3--41.3 & 2.34 & 827.2 & 14.3       & 0.91 & 1259.1$^1$ & 17.7$^1$ \\
        41.3--66.3 & 2.74 & 851.3 & 11.1       & 1.11 & 1259.1$^1$ & 17.7$^1$ \\
        66.3-99.1 & 3.65 & 851.3 & 20.4       & 2.27 & 1259.1 & 17.7 \\
        134.8-164.0 & 1.90 & 846.6 & 12.9       & 0.38 & 1244.9  & 18.8$^2$ \\
      \bottomrule
   \end{tabular}
   \caption{Parameters of the two \Co lines as fitted in four epochs shown in Fig.~\ref{fig:SPI-spectra-set-847-1238}. The time tag is given as the center of the epoch, in days after explosion. Fluxes are given in units of 10$^{-4}$~ph~cm$^{-2}$s$^{-1}$, line centroid and width as Gaussian width $\sigma$ in keV units. Annotations: $^{(1)}$ Value fixed to value fitted in 3rd epoch; $^{(2)}$ Value fixed to value fitted for 847 keV line.}
   \label{table:56Coline-parameters}
\end{table}

At late epoch (4), we find the expected pair of clear (at least at 847~keV) and broadened lines near the rest energies at 847 and 1238 keV, with a broadening of \about~30 and 44~keV (FWHM), respectively. This broadening is equivalent to a velocity spread of \about~(4570$\pm$1840)~km$^{-1}$, determined from the 847~keV line.
From Fig.~\ref{fig:SPI-spectra-set-847-1238}, it is apparent that the 847 keV line varies in position and width among the four different epochs. Also apparent is that there may be multiple emission features in the 780--920~keV energy band (left column in Fig.~\ref{fig:SPI-spectra-set-847-1238}).
We characterise the spectra through Gaussians on top of a continuum offset, fitting the main, broadened, line features in the spectra plus a second, narrow, Gaussian in the 780--920~keV range.  

Tracing the signature for the 847~keV line from the late epoch (4) towards earlier epochs, we can identify a consistent, broadened line, slightly blue-shifted and broader in epochs (3) and (2), while  in epoch (1) the broad feature appears red-shifted towards 827~keV centroid energy, i.e. by \about~6,920~km$^{-1}$. 
We chose to identify this broad feature in the spectra representing the emission of the bulk of \Co from the 847~keV line, but we note some arbitrariness here in particular for the first epoch. If we chose to identify the narrow line feature near 860~keV with 847~keV \Co, we would obtain the expected blue shift at early time, but the line width, and the origin of the emission centered at 827~keV, would be puzzling.  
The fitted parameters for the broad lines are shown in Table~\ref{table:56Coline-parameters}.

The 1238~keV line appears with a centroid at (1245$\pm$5)~keV in the last epoch (4).
The generally-weaker intensity levels for this line do not allow an independent determination of line shape parameters.  We therefore fix the location and width for the early epochs (1) and (2) with their weak signal on the values fitted in epoch (3), and only fit line intensities here. For the weaker signal in epoch (4), we chose to fix the 1238~keV line width to the Doppler broadening determined for the 847~keV line, and only fit intensity and centroid (we assume that the supernova is transparent to \Co gamma-rays here and thus both lines should reflect all \Co from the supernova in the same way). We thus allow to some extent for different bulk velocities of the observed \Co decay at the four epochs, trying to derive as much as we can from the data themselves. 
We find the 1238~keV line somewhat blue-shifted earlier in (brighter) epochs (2) to (3) at 1259~keV; statistical precision is inadequate to determine its variation at different epochs, in particular the red-shift as indicated in epoch (1) for the 847~keV line cannot be constrained nor confirmed independently. 

% Table of Line fitting derived results
\begin{table}[htbp]
   \centering \small
   \begin{tabular}{@{} ccc @{}} % Column formatting, @{} suppresses leading/trailing space
      \toprule
   %  \multicolumn{4}  {c}  {\quad\quad 847 keV line}   {1238 keV line
    % } \\
      %\cmidrule(r){2-3} % Partial rule. (r) trims the line a little bit on the right; (l) & (lr) also possible
      %\cmidrule(l){4} % Partial rule. (r) trims the line a little bit on the right; (l) & (lr) also possible
    
      time  & bulk & spread       \\
      \midrule
        16.3--41.3    & -6920$\pm$1480 & 5060$\pm$1330   \\
        41.3--66.3    & 1600$\pm$1720  & 3940$\pm$1260   \\
        66.3-99.1     & 1600$\pm$1600  & 7250$\pm$1560   \\
        134.8-164.0 & -80$\pm$1870    & 4570$\pm$1840   \\
      \bottomrule
   \end{tabular}
   \caption{Velocity values in km~s$^{-1}$ as derived from the 847~keV \Co line fits in four epochs shown in Fig.~\ref{fig:SPI-spectra-set-847-1238}, and listed in Table~\ref{table:56Coline-parameters}. The 1238~keV line derived centroid and spread are only determined in brightest epoch 3, and are  5050$\pm$1240 and 4290$\pm$1370 km~s$^{-1}$. Fixing the 1238 keV line's width to the value determined from the 847 keV line, its centroid also for the weak epoch (4) can be found, at 1610$\pm$960~km~s$^{-1}$. }
   \label{table:56Coline-velocities}
\end{table}

Table~\ref{table:56Coline-velocities} summarises the velocity constraints from this four-epoch analysis. Overall, both lines are consistent in their centroids and broadenings, within uncertainties, although differences are remarkable.
 {The bulk Doppler shifts  differ somewhat between the two lines, even at late times (near transparency) we find -80$\pm$1870~km~s$^{-1}$ and 1610$\pm$960~km~s$^{-1}$ for the 847 and 1238 keV lines, respectively. This may tell us that the \Co visible in each of the lines reflects a different subset of the total, with different spatial sampling and thus kinematics. }
Note that at early times, it is not clear \emph{a priori} which fraction of the \Co of the 3-dimensional exploding supernova is visible, and what the magnitude of occultation is, as both \Ni and overlying ejecta morphologies are unknown, while towards late times, the true kinematic signature of all of the \Ni should be reflected in both \Co lines in a consistent way. 
The transparency of the supernova envelope is expected to vary with energy, transparency being higher at higher gamma-ray energy.
 
%{\it NOTE: this is not a good explanation for the inconsistency of the late eopoch (4) values, which both should be the same, and with essentially zero shift for an overall-symmetric explosion! But we may argue that for the 'transparent' epoch (4), the 1238 keV line signal is so weak that a line at 1238 keV with the broadening seen in the 847 keV line actually would be a good / acceptable fit, and thus that the brighter epochs (2) and (3) show actually a blue-shifted 1238 keV line, and epoch (1) shows two lines, one the same blue-shifted one, and the other showing a red-shift consistent with what we see in the 847 keV line. We should provide a quantitative assessment of testing this hypothesis, based on the 847 keV line, testing the 1238 band for a consistent result or not.}

% ----------------------------------------------------------------------
\begin{figure}
  \centering
  \includegraphics[width=\linewidth]{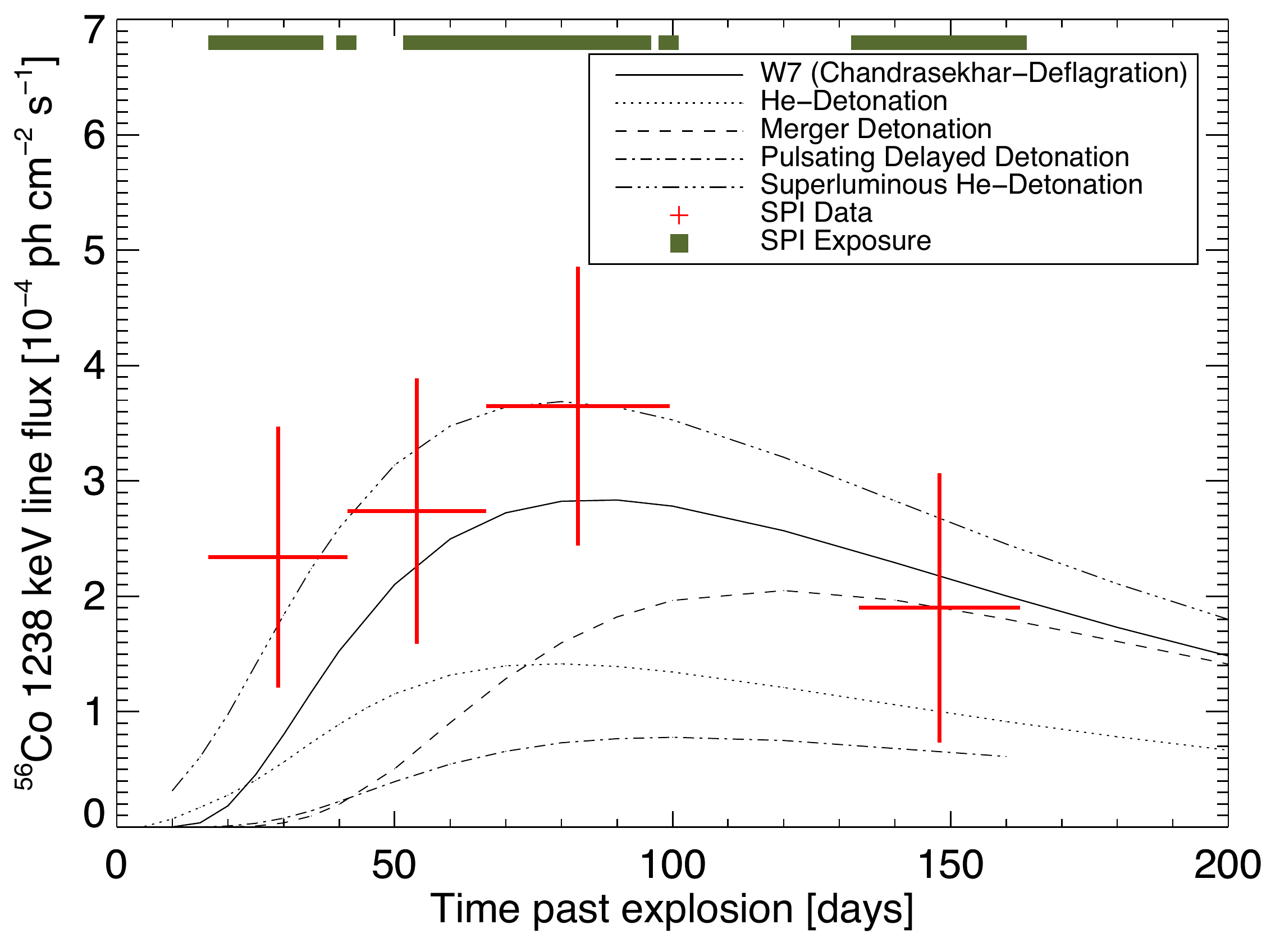}
  \includegraphics[width=\linewidth]{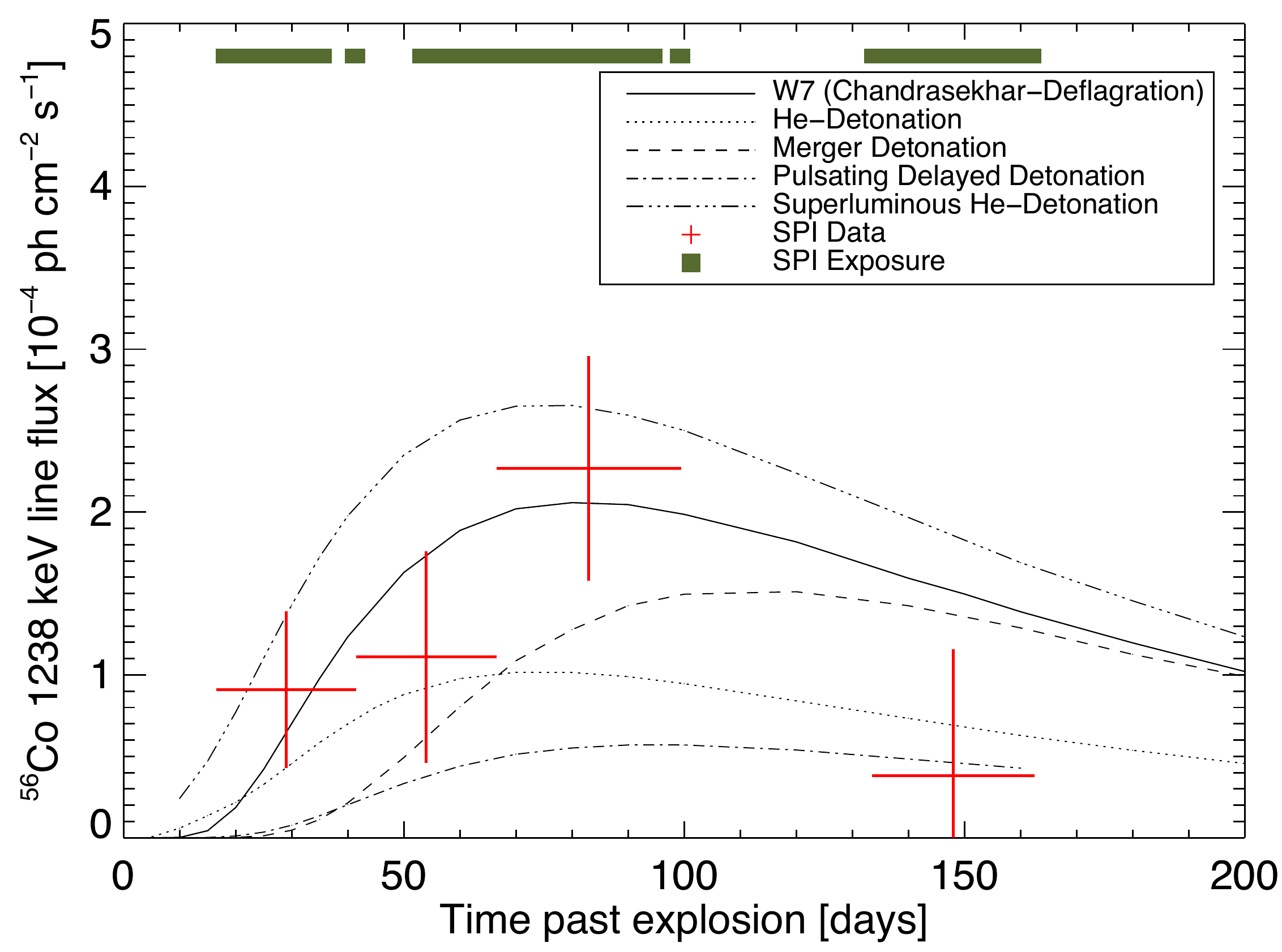}
   \caption{SN2014J signal intensity variations for the 847 keV line ({\it above}) and  the 1238 keV line ({\it below}) as seen in our observations and shown in Fig.~\ref{fig:SPI-spectra-set-847-1238}. Here the intensity was derived by Gaussians fitted to the spectra near the respective \Co line at four independent epochs, as discussed in the text. The epochs are shown as horizontal bars at each data point. For reference, several candidate model light curves are shown, as extracted from \citet{The:2014}.
  }
  \label{fig:SPI-lightcurve-847-1238}
\end{figure}
% ----------------------------------------------------------------------

The intensity variation throughout these four epochs of our observations for both lines produce gamma-ray light curves shown in Fig.~\ref{fig:SPI-lightcurve-847-1238}. 
Evidently, both lines consistently rise towards a maximum near 60--100~days after explosion, falling off later. The intensity ratios between both lines (see Table~\ref{table:56Coline-parameters}) generally agree (within uncertainties) with the nominal branching ratio of \Co decay, in particular for the brightest epoch (3) we find the 1238 keV line being at 62\% ($\pm$28\%) of the 847 keV line intensity, which compares to a laboratory value of 68\%. 
%But the observed variation provides a hint that the more-penetrating higher-energy line may be observed from a slightly different volume of \Ni than the lower-energy line. 

\begin{figure}
  \centering
  \includegraphics[width=\linewidth]{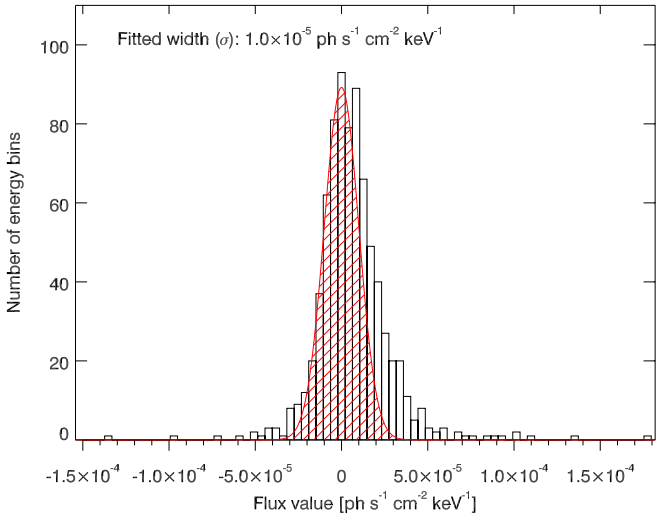}
   \caption{SN2014J flux histogram in the 780--920 keV band including the 847 keV line. The histogram of flux values shows an excess at positive flux values above statistical fluctuations with a significance of 11.3~$\sigma$ (details see text).}
  \label{fig:SPI-fluxhist-847}
\end{figure}

For assessment the detection of supernova emission from \Co decay, we also check upon our statistical uncertainties.
%We note that we fit the Gaussians on top of an offset or baseline flux, but some datapoints formally show negative flux values, although none of those values is significantly below zero. 
For an independent estimate of uncertainties, we histogram the resulting SN2014J flux values, and compare their distribution near zero flux with expectations from Poissonian statistical uncertainty (these are the error bars shown in Fig.~\ref{fig:SPI-spectrum-847-1}). 
Fig.~\ref{fig:SPI-fluxhist-847} shows the flux value histogram for all observations in the 780--920~keV band, analysed in the above eleven time bins, and 2~keV energy bins. The distribution of values towards negative fluxes from zero is due to statistical fluctuations only, while a celestial source contributes to the distribution at positive flux values. 
We thus obtain flux uncertainties from the negative part of the distribution, and total source significance from its  positive part. This yields a KS-test p-value of 1.4~10$^{-29}$ for the positive flux values following the same statistics-only distribution, which is equivalent to a probability of 11.3~$\sigma$ that the spectrum contains nonzero signal from SN2014J. 
Measuring the width of the distribution of fluxes below zero, we find 1.0~10$^{-5}$~ph~cm$^{-2}$s$^{-1}$, which compares to a Poissonian error of $\pm$0.54~10$^{-5}$, i.e. 1.08~10$^{-5}$~ ph~cm$^{-2}$s$^{-1}$ (the error in data points shown in the spectra of Fig.~\ref{fig:SPI-spectrum-847-1}).  
We conclude that our Poissonian statistical error estimates are approximately correct.
Using this assessed statistical uncertainty, the narrow-line signals from SN2014J shown in Fig.~\ref{fig:SPI-spectra-set-847-1238} in the 780--920~keV energy range correspond to a statistical significance of \about~2.8--4~$\sigma$.
%, together with the spectrum for all observations shown in Fig.~\ref{fig:SPI-spectrum-847-1}. 
%We use these results also to independently estimate the magnitude of statistical fluctuations and thus the overall significance of our signal from SN2014J.
%The flux histograms of spectra in the energy band 780 -- 920 keV are shown in Fig.~\ref{fig:SPI-fluxhist-847} for energy bins of 10 and 1 keV. The 1~keV energy bin spectrum shown in Fig.~\ref{fig:SPI-spectrum-847-1}, clearly, shows more statistical noise, as SN2014J signal per energy bin is reduced. {\it Add here the 1 keV histogram and significance description, and a statement on the 1238 keV histograms.}

% ----------------------------------------------------------------------
% ----------------------------------------------------------------------

\subsubsection{Exploring shorter time variations and high spectral resolution}

The spectrum in the 847~keV line band of SN2014J (Fig.~\ref{fig:SPI-spectra-set-847-1238}) shows, beyond the broadened line discussed above, that apparently additional but narrow lines seem to be present at significance levels exceeding 3~$\sigma$, near 865~keV in epoch (1), 805~keV in epoch(2), and 815~keV in epoch (3). 
No such features are apparent in the high-energy band 1190--1290~keV.

The large instrumental background may be a concern, creating artificial transient features. 
We verified adequacy and consistency of our background modeling in these epochs. The orbit-by-orbit spectral fits do not show indications for deviations of fit quality in those energy regions and times. 
None of the background components show correlations with the sky signal.
 
 % {\it (Discuss line-fitting uncertainty (MCMC simulations of fitting trials?), difficulties to assign lines to isotopes. Check Kromer's spectral features for other candidate lines.)} 

% Table of Line fitting results - different line shapes
\begin{table*}[htbp]
   \centering \small
   \begin{tabular}{@{} cccr @{}} % Column formatting, @{} suppresses leading/trailing space
      \toprule
Epoch  &  $\chi^{2}_{best}$ (dof)  &  $\chi^{2}_{alltimes}$ ($\sigma$)  &  $\chi^{2}_{from4thepoch}$ ($\sigma$) \\
    \midrule
1      &  12.62 (6)                &  42.05 (4.28)                      &  42.93 (4.36)                         \\
2      &  17.77 (6)                &  27.48 (1.73)                      &  29.35 (2.04)                         \\
3      &   5.77 (6)                &  15.37 (1.71)                      &  15.79 (1.78)                         \\
4      &  24.78 (9)                &  25.54 (0.41)                      &  - (-)                                \\
     \bottomrule
   \end{tabular}
   \quad\quad
   \begin{tabular}{@{} cccr @{}} % Column formatting, @{} suppresses leading/trailing space
      \toprule
Epoch  &  $\chi^{2}_{best}$ (dof)  &  $\chi^{2}_{alltimes}$ ($\sigma$)  &  $\chi^{2}_{from4thepoch}$ ($\sigma$) \\
    \midrule
1      &  48.36 (62)               &  66.85 (3.04)                      &  66.82 (3.03)                         \\
2      &  66.90 (62)               &  77.73 (1.92)                      &  76.58 (1.72)                         \\
3      &  43.22 (62)               &  60.95 (2.94)                      &  58.81 (2.65)                         \\
4      &  50.66 (65)               &  51.18 (0.29)                      &  - (-)                                \\
     \bottomrule
   \end{tabular}
  \caption{Line shape variations of the 847 keV line. For each of the four observation epochs, we list the $\chi^2$-values of the best fit (column 2; Gaussian with adapted centroid and width), then the intensity-fitted shape of the line as determined in the time-integrated results (Fig.~\ref{fig:SPI-spectrum-847-1}) (column 3), and finally (column 4) the intensity-fitted line shape of the last epoch. The degrees of freedom of the fits are given in column 2 in brackets. The discrepancy of the thus-adopted line shape in each of the epochs is given in $sigma$ units in brackets in columns 3,4. The {\it lefthand} table gives the results for the 10 keV bin data (center column in Fig.~\ref{fig:SPI-lightcurve-847-1238}), the {\it righthand} table gives the results for the 2 keV bin data (left column in Fig.~\ref{fig:SPI-lightcurve-847-1238})} 
   \label{table:lineshapefits}
\end{table*}

Having confirmed the statistical uncertainties (see above), we now investigate how different resolution in time and in spectral bin width affect our analysis results.
Narrow-line features with fluxes at the level of 2~10$^{-4}$~ph~cm$^{-2}$s$^{-1}$ are reminiscent of \Ni decay, because \Ni volume emissivity is much higher due to the shorter decay time than for \Co lines, and thus smaller active volume elements with specific kinematic and Doppler shift might contribute transient lines with a typical 1--2~week time scale. Also, if the \Ni synthesised in the explosion emerges in clumps, each clump may be expected to have different kinematics and somewhat different turbulence, cooling, and expansion.  
We analyse the 780--920~keV band therefore in eleven time epochs independently, and in finer energy bins of 2~keV resolving lines as narrow as the instrumental line width. 
%The resulting set of spectra are shown in Fig.~\ref{fig:SPI-spectra-set-847-1238_11}, with the empty panel indicating the data gap between days 100 and 134 after explosion.
We find that in finer time bins the broad \Co line features are difficult to recognise, except towards the late epoch in days 134--164. We also note that those three late-epoch spectra are otherwise featureless and rather flat, which is reassuring. 
However, all earlier spectra show, more or less, spectral features which are less broad, and which altogether combine to form the broad features seen at coarser energy binning. It appears that between 830 and 860~keV, \about~10~keV wide features emerge at times before day 99, but vary in intensity and peak location detail, with a consistent trend towards lower energies. Moreover, around 810~keV, another line feature emerges before days 88, which started to emerge only after days \about~60. The feature near 865~keV apparent in the first sample is gone after days 25, while near 820~keV, emission is indicated near days 30 and 55, not seen in adjacent epochs. 
We conclude %, that the narrower (5--10~keV wide) line-like features that we may see in \about~10-day samples are transient. 
that, while being consistent with appearance of the 847~keV line emission, this variability is surprising. { When we assume an intrinsic line shape of \Co as seen at late times, or integrated and as seen in broad energy bins (Fig.~\ref{fig:SPI-spectrum-847-1}), we find that the line shapes in the different epochs shown in Fig.~\ref{fig:SPI-lightcurve-847-1238} are significantly different in each case (see Table~\ref{table:lineshapefits}). This may be understood as different sight lines to embedded \Co determining the observed lines at each of our epochs, which in turn implies that the envelope of the supernova includes 3D structure which evolved during the time of our gamma-ray observations. }  

\Ni decay has two strong lines, one at 812~keV, and another one at 158~keV line. We find no  emission near 158~keV coincident with the narrow line features in the 800-900 keV band, and therefore we can exclude an origin in late \Ni emission (e.g. from fully-ionised \Ni regions). 

If we take our results as showing three distinct emission components, we may employ a simple model to treat them individually in terms of \Ni content, kinematics, and occultation, embedded in an expanding supernova. We roughly identify components at 810$\pm$5 keV, 845$\pm$10 keV, and  865$\pm$7.5 keV. We find that flux time histories of each component are consistent with early occulted and later revealed decaying \Co line emission, and we determine optical depths at day 1 of 2000, 600, and 100 for the 810, 845, and 865~keV centered components, respectively. (High opacity for the volume emitting 810 keV \Co arises due to early contributions from \Ni in the 812 keV line.) The kinematics suggested for such three identified major \Co clumps are roughly blue-shifted 6400~km~s$^{-1}$, (\about~15\% of the \Ni), red-shifted 13,000~km~s$^{-1}$, (\about~30\% of \Ni), and a main, spherically symmetric contribution (\about~55\% of \Ni). While we do not propose this to be the real morphology of SN2014J, this analysis supports our interpretation of aspherical \Ni distribution and differences in their respective occulting outer supernova material.

In conclusion, we speculate that occultation and its evolution as the supernova expands may be responsible for the apparent spectral signatures. Clumps or co-moving volume elements carrying \Co may lie along a less-occulted line of sight at specific times, thus contributing emission in a particular bulk Doppler shifted energy regime. As the supernova expands, different volume elements may thus contribute at different times, as long as occultation is significant. This may reflect asymmetry in SN2014J, which had been discussed earlier to characterise a subset of SNIa where multiple or staged explosions occur \citep{Maeda:2010an,Maeda:2010a}.

% ----------------------------------------------------------------------
\begin{figure}
  \centering
  \includegraphics[width=\linewidth]{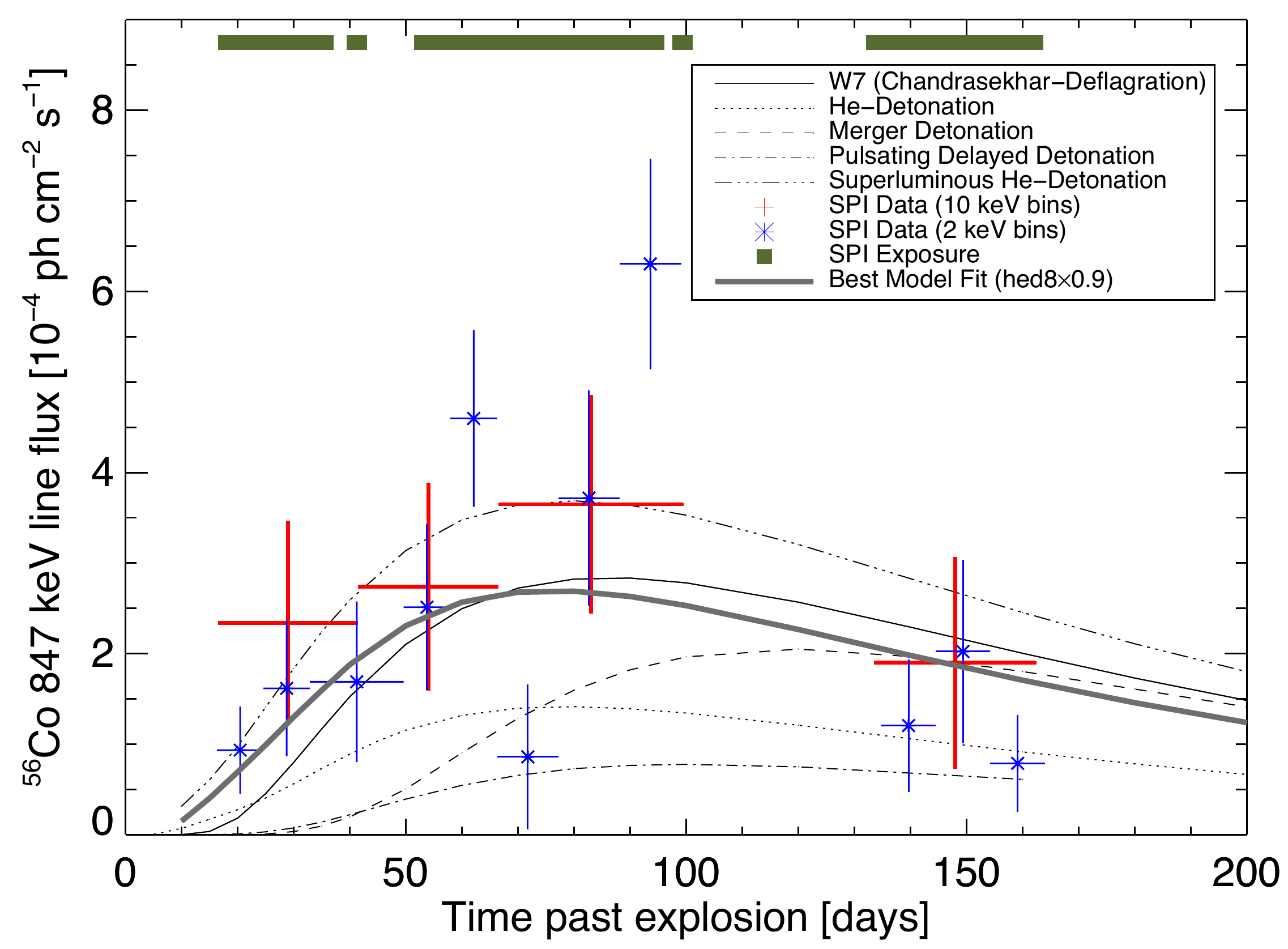}
   \caption{SN2014J signal intensity variations for the 847 keV line in two different time resolutions. The 4-epoch results are consistent with 11-epoch analysis, both showing an initial rise and late decline of \Co decay line intensity, with a maximum at 60--100 days after explosion. Shown are also several candidate source models from  \citet{The:2014}, fitted in intensity and thus determining \Ni masses for each such model. The best-fitting model is shown as a continuous thick line. }
  \label{fig:SPI-lightcurve-847-fine}
\end{figure}
% ----------------------------------------------------------------------

\section{Model comparison and \Ni mass}

% Table of Line fitting results
\begin{table*}[htbp]
   \centering \small
   \begin{tabular}{@{} lccccccccr @{}} % Column formatting, @{} suppresses leading/trailing space
      \toprule
  \multicolumn{3}  {c} \quad line fit    &	& & 100 keV band  & & &	20 keV band \\
    \midrule
 rank     &   model         & $\chi^2$ & M(\Ni)	&   model & $\chi^2$ & M(\Ni) &   model & $\chi^2$ & M(\Ni) \\
      \midrule
1.          &    hed8	   &  25.07   &	0.46$\pm$0.06	& hed8  &	7.81	& 0.52$\pm$0.12 & W7 & 16.98 & 	0.42$\pm$0.07 \\
2.          &	w7dt	           &  25.37  & 0.50$\pm$0.06    & w7dt    & 7.89	    & 0.57$\pm$0.13    & det2    & 17.07    & 0.38$\pm$0.06 \\
3.          &	W7A	          &   25.57  & 0.51$\pm$0.06    & W7A    & 7.95    & 0.59$\pm$0.13    & m36    & 17.15    & 0.43$\pm$0.07 \\
4.          &	hecd	         & 25.61  & 0.52$\pm$0.06    & hecd    & 8.11    & 0.58$\pm$0.13    & dd202c    & 17.26    & 0.42$\pm$0.07 \\
5.          &	hed6	         & 26.22  & 0.50$\pm$0.06    & hed6    & 8.54    & 0.56$\pm$0.13    & W7E    & 17.29    & 0.43$\pm$0.07 \\
6.          &	det2	         & 27.84  & 0.50$\pm$0.06    & det2    & 9.35    & 0.55$\pm$0.13    & dd4    & 17.33    & 0.45$\pm$0.07 \\
7.          &	w7dn	& 28.27  & 0.54$\pm$0.07	    & w7dn    & 9.57    & 0.59$\pm$0.14    & w7dn    & 17.33    & 0.40$\pm$0.06 \\
8.          &	W7E	        & 29.48  & 0.56$\pm$0.07	    & W7E    & 10.11    & 0.61$\pm$0.15    & det2e2    & 17.67    & 0.44$\pm$0.07 \\
9.          &	m36	        & 30.30  & 0.56$\pm$0.07	    & m36    & 10.44    & 0.61$\pm$0.15    & hed6    & 17.73    & 0.37$\pm$0.06 \\
10.        & 	W7	        & 30.52  & 0.54$\pm$0.07	    & W7    & 10.50	& 0.59$\pm$0.14    & hecd    & 18.13    & 0.38$\pm$0.06 \\
11.        & 	dd202c    & 30.70  & 0.55$\pm$0.07    & dd202c    & 10.51    & 0.61$\pm$0.15    & hed8c    & 18.17    & 0.33$\pm$0.05 \\
12.        & 	dd4         & 32.72  & 0.57$\pm$0.08    & dd4    & 11.35    & 0.62$\pm$0.15    & w7dt    & 18.33    & 0.36$\pm$0.06 \\
13.        & 	det2e2     & 34.18  & 0.56$\pm$0.07    & det2e2    & 11.80	     & 0.61$\pm$0.15    & W7A    & 18.40    & 0.38$\pm$0.06 \\
14.        & 	pdd54      & 38.84  & 0.56$\pm$0.08    & pdd54    & 13.52    & 0.61$\pm$0.16    & pdd14    & 18.74    & 0.46$\pm$0.07 \\
15.        & 	det2e6     & 50.75  & 0.72$\pm$0.11    & det2e6    & 17.80	    & 0.78$\pm$0.25    & det2e6    & 22.94    & 0.66$\pm$0.11 \\							
             &	constant      & 35.20  & 0.30$\pm$0.06    & constant    & 8.48    & 0.41$\pm$0.12    & constant    & 26.86    & 0.24$\pm$0.06 \\
             &	bgd only     & 90.92  & -    & bgd only &	27.87    & -    & bgd only    & 57.33    & - \\
     \bottomrule
   \end{tabular}
   \caption{\Ni mass fit results for different explosion models, as described in detail by \citet{The:2014}. { We fit the light curves resulting from those models to our gamma-ray data. The fit} quality can be seen from the $\chi^2$ values { (for 9 degrees of freedom, except for the bgd-only case which has 10 d.o.f.)}. Fitted \Ni masses are given in \Msol with uncertainties, for the intensity variation time profile of the respective model. We compare results for three analysis approaches, fitting a single broad line \emph{(left)}, for taking the non-zero flux in a broad band \emph{(middle)} (800--900~keV), and for taking nonzero flux in a band where the \Co emission could be expected to appear \emph{(right)} (840--860~keV).} 
   \label{table:model-parameters}
\end{table*}

We may compare the time evolution of our measured fluxes of \Co line emission to expectations from different models. For this, we use standard models derived for generic plausible assumptions over the past decades, as presented and discussed by, e.g., \citet{Nomoto:1984aa,Khokhlov:2001aa,Woosley:1994aa,Livne:1995aa,Hillebrandt:2000aa,Mazzali:2007aa,Isern:2011ac,Milne:2004aa,Hillebrandt:2013aa,Dessart:2014}. Recenlty, \citet{The:2014} have recompiled a representative set of models specifically for applications to SN2014J, and we use the set discussed there in our aim to capture key properties that may describe SN2014J best. % (the 4-epoch results are also included for comparison). 

We return to the issue of how to properly assign \Co line emission to earlier epochs, given that the observed emission is not represented by a broad (Gaussian) shape, but rather by evolving appearance of emission across the energy range finally described through such a broadened line (see Fig.~\ref{fig:SPI-spectra-set-847-1238}); we also remind that Comptonisation will incur some apparent redshift of \Co line emission, while the kinematics of \Ni ejected in all directions will imprint its own Doppler shifts, and will be revealed as the overlying supernova material becomes increasingly transparent. 
This implies that the spectral shape of \Co emission is uncertain and depends on the 3-dimensional distributions of \Ni and of occulting, overlying material, which is not accounted for in any of these (1-dimensional) models. 
Therefore, we concentrate on the brighter, more-clearly measured \Co line at 847~keV, and then we compare three different approaches to determine \Co decay emission originating from the 847~keV line in our different epochs: 

Approach (a) uses the line parameters as fitted in late epochs, and fits earlier epochs constraining line width and centroid to within the 2$\sigma$ uncertainty limits of that late-epoch fit. Thus we stabilise the way we infer line emission from this energy region, using the late-epoch emission as the constraint, within earlier revealed emission spots must fall. Model-fitted \Ni masses are in the range 0.46--0.52~($\pm$0.06)~\Msol for most-plausible models with some \Ni near the supernova surface (see  \citet{The:2014}). Another approach (b) assumes, as another extreme, that \emph{all} emission above the zero level in the 800--900~keV energy band may be attributed to \Co emission originating from  the 847~keV line emission.
%, producing results shown in Fig.~\ref{}. 
Clearly, this over-estimates early epoch intensties, which, additionally, may carry contributions from the more-intense \Ni decay. Correspondingly, model-fitted \Ni masses are higher and in the range 0.52--0.59~($\pm$0.13)~\Msol. Approach (c) is an intermediate case, evaluating the energy range of the broad 847~keV line, yet not prescribing a spectral shape constraint. Averaging over the three best-fitting models in our three approaches, we derive a \Ni mass of 0.49$\pm$0.09~\Msol.

{We adopt the model-predicted \Ni masses and \Co line fluxes versus time for each of the models, and fit those \emph{light curves} to our data as defined in approaches (a) to (c), scaling their intensity. This scaling factor then determines the \Ni mass for SN2014J from each model, as fitted to our observations.}
The results shown in Table~\ref{table:model-parameters} provide SN2014J \Ni masses in the range \about~0.4 to 0.8 \Msol. The best-fitting models with their He on the outside also probably give the best \Ni mass estimates around  (0.5$\pm$0.1)~\Msol. The time profiles of even the best-fitting models do not match in detail the fluxes determined for a single broad line, but improve significantly when the spectral constraints are relaxed, thus indirectly confirming our above claim for irregular appearance of \Co lines. In Table~\ref{table:model-parameters} we also see that our observations, while providing a clear detection of the \Co decay gamma-ray lines, are still inadequate to clearly discriminate among models; most models are in agreement with our observed intensity evolution of the gamma-ray lines, if \Ni is a free parameter. Better constraints could be expected, if models with different \Ni and ejecta morphologies and corresponding radiation transport would be available, and tested on the broader energy range including all \Co lines and continuum emission as it may evolve.

%%%%%%%%%%%%%%%%%%%%%%%%%%%%

\section{Conclusions}
\label{sect-conclusions}

We analyse the full set of INTEGRAL/SPI observations available for SN2014J, and concentrate on the energy ranges around the brightest gamma-ray line emission expected from \Co decay, around the two bright lines at 847 and 1238~keV energy. We employ an analysis that uses the coded-mask shadowgrams as the source aspect is varied across those observations, and searches for emission consistent with these variations, above a large instrumental background.
The instrumental background is derived by a detailed spectral analysis and modeling, which accounts for continuum and many instrumental lines arising from cosmic-ray interactions with instrument and spacecraft, in a broader energy range around the lines of interest. In this, we first identify long-time spectral features, then separately fit their appearance in each individual Ge detector with its response characteristics, and finally determine the amplitude variations of the background spectral model on short time scales of our individual pointings. The model background is then extracted for the energy bins where we aim to determine a flux determination from SN2014J. This method overcomes limitations of statistical precision of background measurements in the fine energy bins where we wish to analyse the source, and uses data as measured by the SPI detectors themselves, and the physics extracted from longterm monitoring of background and detector responses.  

Our  measurements of  \Co decaying in SN2014J find the two strongest gamma-ray lines at 847 and 1238 keV, clearly distinguished from instrumental background and attributed to the supernova. We analyse separately different epochs of our observations, and derive an intensity time profile in the energy regime around these two lines. The brightness evolution is consistent {between} both lines, and is overall consistent with expectations from different plausible explosion models. {We find that models preferred by the observed gamma-ray light curve also include those with \Ni on the outside. We obtain \Ni mass values of 0.5--0.6~\Msol for best-fitting models, here determined from gamma-ray light curve data of the \Co decay for the first time.} We note, however, that the appearance of the line emission is irregular and not recognised as a smooth emergence of the Doppler broadened emission lines as they are found in the late, gamma-ray transparent observations of SN2014J. We find transient emission features of different widths, but cannot assign them clearly to origins of \Ni and \Co at specific kinematic properties. We infer that the 3-dimensional structure of the inner supernova, and possibly also of the material  overlying the \Ni, are not as regular as 1-dimensional models of gamma-ray emergence currently implement.
  
Our data cannot discriminate among models, although Chandrasekhar-mass models are slightly less favoured. 
%The observed line positions and widths suggest that the explosion may not be spherical, and  ejected envelope and/or \Ni morphologies could be complex or clumpy. Although these results indicate the potential of gamma-ray line observations, SN2014J was still not bright (or close) enough for the current generation gamma-ray telescope such as operating on INTEGRAL.  
The observed line shapes clearly suggest that the explosion was not spherically symmetric. 
These results demonstrate the power of gamma-ray line observations from type Ia supernovae, and seed high expectations should a closer event occur during the INTEGRAL lifetime.

% ----------------------------------------------------------------------
%
\begin{acknowledgements}
  This research was supported by the German DFG cluster of excellence 'Origin and Structure of the Universe', and from DFG Transregio Project No. 33 'Dark Universe'. F.K.R. was supported by the DFG (Emmy Noether Programm RO3676/1-1) and the ARCHES prize of the German Ministry for Education and Research (BMBF). The work by K.M. is partly supported by JSPS Grant-in-Aid for Scientific Research (23740141, 26800100) and by WPI initiative, MEXT, Japan. W.W. is supported by the Chinese Academy of Sciences and the Max Planck Gesellschaft, and by the German DFG cluster of excellence 'Origin and Structure of the Universe'. We are grateful to E. Kuulkers for handling the observations for the INTEGRAL SN2014J campaign.
  The INTEGRAL/SPI project
  has been completed under the responsibility and leadership of CNES;
  we are grateful to ASI, CEA, CNES, DLR, ESA, INTA, NASA and OSTC for
  support of this ESA space science mission.
  INTEGRAL's data archive (http://www.isdc.unige.ch/integral/archive\#DataRelease) is at the ISDC in Versoix, CH, and includes the SN2014J data used in this paper.
\end{acknowledgements}
% ----------------------------------------------------------------------
%
\bibliographystyle{aa}
%\bibliography{rod-refs_SNIa,rod-refs13}

%
% ----------------------------------------------------------------------

%\appendix
%\section{Spectra at high resolution and short epochs}

% ----------------------------------------------------------------------
%\begin{figure*}
%  \sidecaption
%  \includegraphics[width=12cm]{Fig_11epochs_spec_847_1keV}
%   \caption{SN2014J signal intensity variations for the 847 keV line as determined for eleven epochs of our observations, in fine (instrumental-width) energy resolution. The same signal already shown above in Fig.~\ref{fig:SPI-spectra-set-847-1238}, can be inferred also in these different time samplings, and refined energy resolution, although increased statistical fluctuations reduce the S/N. Narrow features are recognized over a wider energy range in the earlier observations, settling to a more quiet spectrum at late epochs, where the broad emission from \Co decay can be recognized.}
%  \label{fig:SPI-spectra-set-847_11}
%\end{figure*}
% ----------------------------------------------------------------------

\end{document}